\documentclass[fleqn,twocolumn,10pt]{wlscirep}
\usepackage[utf8]{inputenc}
\usepackage[T1]{fontenc}
\usepackage{hhline}
\usepackage{amsmath}
\usepackage{multirow}
\usepackage{subfig}
\usepackage[absolute]{textpos}
\title{Advanced sleep spindle identification with neural networks}
\author[1]{Lars Kaulen}
\author[2]{Justus T.~C.~Schwabedal}
\author[3]{Jules Schneider}
\author[3]{Philipp Ritter}
\author[1,4,*]{Stephan Bialonski}
\affil[1]{Department of Medical Engineering and Technomathematics, FH Aachen University of Applied Sciences, 52428 Jülich, Germany}
\affil[2]{Independent researcher, Lessingstraße 65, 53113 Bonn, Germany}
\affil[3]{Department of Psychiatry and Psychotherapy, University Hospital Carl Gustav Carus, Technische Universität Dresden, 01307 Dresden, Germany}
\affil[4]{Institute for Data-Driven Technologies, FH Aachen University of Applied Sciences, 52428 Jülich, Germany}
\affil[*]{bialonski@fh-aachen.de}

\newcommand{\Fmean}{$\overline{\text{F1}}$}

\begin{abstract}
Sleep spindles are neurophysiological phenomena that appear to be linked to
memory formation and other functions of the central nervous system, and that can be observed
in electroencephalographic recordings (EEG) during sleep.
Manually identified spindle annotations in EEG recordings suffer from
substantial intra- and inter-rater variability, even if raters have been highly
trained, which reduces the reliability of spindle measures as a research and
diagnostic tool.
The Massive Online Data Annotation (MODA) project has recently addressed this
problem by forming a consensus from multiple such rating experts, thus providing
a corpus of spindle annotations of enhanced quality.
Based on this dataset, we present a U-Net-type deep neural network model to
automatically detect sleep spindles.
Our model's performance exceeds that of the state-of-the-art detector
and of most experts in the MODA dataset.
We observed improved detection accuracy in subjects of all ages, including older
individuals whose spindles are particularly challenging to detect reliably.
Our results underline the potential of automated methods to do repetitive
cumbersome tasks with super-human performance.
\end{abstract}

\begin{document}

\flushbottom
\maketitle

\begin{textblock*}{21cm}(0cm,26.5cm)
  \centering
  This is a preprint of an article published in Scientific Reports 12, 7686 (2022).\\
  The final authenticated version is available online at: \href{https://doi.org/10.1038/s41598-022-11210-y}{https://doi.org/10.1038/s41598-022-11210-y}
\end{textblock*}

\thispagestyle{empty}

\section{Introduction}
\label{sec:intro}

Sleep spindles are phenomena occurring in the brain of sleeping mammals
\cite{Fernandez2020} that are associated with diverse neurological processes,
such as cognitive abilities, including learning, memory consolidation and
intelligence\cite{Schabus2006,Fogel2011,Rasch2013,Lustenberger2015,Fernandez2020}.
Changes in sleep spindle characteristics have been
reported for various sleep disorders\cite{Normand2016}, mental illnesses such
as schizophrenia\cite{Ferrarelli2007, Wamsley2012, Manoach2016}, autism
\cite{Limoges2005, Tessier2015} and bipolar disorder\cite{Ritter2018}, and
neurodegenerative diseases such as Alzheimer's disease\cite{Petit2004,
Weng2020}. These findings suggest that sleep spindles may serve as
prognostic and diagnostic biomarkers for various diseases in the future
\cite{Fernandez2020}.

Spindles are brief, distinct bursts of rhythmic activity visible in
electroencephalographic recordings (EEG) with frequencies in the range of
11--16\,Hz (sigma-band) and a duration of at least $0.5$ seconds, according to
the American Academy of Sleep Medicine (AASM)\cite{Berry2018}.
Sleep spindles are generated in the thalamic reticular nucleus and relayed to
the cortex by neuronal feedback loops\cite{Fernandez2020}.
The gold standard
for spindle detection is spindle identification by trained experts who visually
inspect and annotate EEG signals. This manual process is known to be
time-consuming and therefore expensive, and is thus a limiting factor for large-scale
spindle studies. Moreover, inconsistencies and disagreements among
experts introduce noise in spindle annotations (label noise).  Indeed, only substantial
agreement was reported when an expert inspected the same data multiple times
(\textit{intra-rater reliability}, Cohen's $\kappa=0.66$), and the agreement between
multiple experts (\textit{inter-rater reliability}) was only moderate
($\kappa = 0.52$)\cite{Wendt2015}.

An automated system for the detection of spindles that mimics an expert scorer
would determine spindles consistently, but it would be limited by the level of
disagreement among experts.
This challenge can be addressed
by having a group of experts separately identify sleep spindles in order to form a
group consensus (\emph{expert consensus})\cite{Wendt2015}.  By increasing the
number of experts, such a consensus converges to generalizable
annotations\cite{Kraemer1979}.
For sleep spindle identification, it was estimated that 2--3 experts were
required to achieve a consensus of `substantial' reliability, and at least~4~or
more experts were needed to reach `almost perfect' reliability\cite{Wendt2015}.
Publicly available datasets of spindle annotations (such as MASS-SS2 with $19$
participants about $23.6$~years old, see Montreal Archive of Sleep Studies\cite{OReilly2014}) contain recordings of
several subjects that are annotated by only one or two expert scorers.  Notably, the
DREAMS dataset\cite{Devuyst2006} was scored by five expert scorers but only comes from eight
participants that were about $45.9$~years of age\cite{Kulkarni2019}, thus limiting the
generalization of systems trained thereon.
Recently, in a remarkable crowdsourcing effort, the MODA (Massive Online
Data Annotation) dataset was created in which EEG recordings from 180 subjects
were scored for spindles by a median number of 5 experts (more than 95\% of the
data was annotated by at least 3 experts)\cite{Lacourse2020}.
Among the investigated methods for automated spindle identification, the
algorithm A7\cite{Lacourse2019} achieved the highest level agreement with the expert consensus
with a performance similar to an average expert scorer.
However, as the consensus reduces the label noise substantially,
we entertain the hypothesis of an automated system that exceeds individual
expert scorers and replaces the manual annotation process,
thereby enabling large-scale spindle studies.

The algorithms presented in the MODA study, including A7, are based on engineered
features which make use of
well-known physiological and physical properties of sleep spindles as they
appear in the EEG, such as spindle amplitudes and frequencies\cite{Schimicek1994,Moelle2002,Ferrarelli2007,Wamsley2012,
Martin2013,Parekh2015,Lacourse2019}.
The past years have shown that such feature-based methods are inferior to
methods of deep learning with neural networks, especially for problems of
object or event detection\cite{Schmidhuber2015, LeCun2015, Goodfellow2016}.
Recent studies have confirmed this trend by training a variety of neural
network architectures on minimally processed EEG signals: neural-net-based
approaches report superior performance with respect to the F1 score,
a quantity usually evaluated at a 20\% overlap between algorithmically-detected
and expert-annotated spindle.
\textit{DOSED}\cite{Chambon2019} achieved an F1 of about $0.75$, $0.50$, and
$0.45$ when evaluated on a young, middle-aged, and older cohort.
Exceeding these results, \textit{SpindleNet}\cite{Kulkarni2019} achieved an F1
of $0.82$ when trained on a young cohort, but reached only $0.48$ on
DREAMS, whose participants also have a variety of sleep pathologies.
\textit{RED-CWT} and \textit{RED-Time}\cite{Tapia2020}, and
\textit{SpindleU-Net}\cite{You2021} achieved comparable performance of about
$0.83$ on the MASS-SS2 dataset; SpindleU-Net was also evaluated on DREAMS and
exceeded previous results by reaching an F1 of $0.74$.

In this contribution, we introduce a deep neural network model called
\emph{SUMO}\cite{Kaulen2022a} (\emph{S}lim \emph{U}-Net trained on \emph{MO}DA) that approximates the expert consensus derived from the MODA
dataset to identify sleep spindles.  The network architecture is inspired by
U-Net\cite{Ronneberger2015}, a deep neural network that has been used with
great success in the field of image segmentation\cite{Siddique2021}.  The model
complexity of SUMO is low compared to aforementioned deep learning models, which
allows for data-efficient training on the MODA dataset.  We demonstrate that
our model surpasses the accurateness of the A7 algorithm and the average expert
to identify spindles compared to the group consensus.  Downstream measures
(spindle duration and density) derived from the model's predictions correlated
stronger with the expert consensus compared to A7.  We consider approaches such
as ours to be promising for the future of automated sleep spindle
identification.

\section{Materials and methods}
\label{sec:methods}

\begin{table}
	\centering
	
	\begin{tabular}{@{}llrrrrr@{}}
	\toprule
	& & & \multicolumn{2}{c}{test} & \\
	\cmidrule(l){4-5}
	& cohort & training & younger & older & sum \\
	\midrule
	individuals & younger & 82 & 18 & 0 & 100 \\
	& older & 62 & 0 & 18 & 80 \\
	& sum & 144 & 18 & 18 & 180 \\
	\midrule
	segments & younger & 351 & 54 & 0 & 405 \\
	& older & 291 & 0 & 54 & 345 \\
	& sum & 642 & 54 & 54 & 750 \\
	\bottomrule
	\end{tabular}
	\caption{
	Training- and test-splits of individuals and EEG segments, respectively.
	We divided all individuals in the MODA dataset into a training and test set
	such that, for the test set, individuals and segments were divided equally between age groups.
	Among a set of random splits satisfying this equality, we selected the split with
	median performance for algorithm A7.
	\label{tab:data:splits}
	}
\end{table}
	
\subsection{EEG and spindle dataset}

We obtained two datasets to train and study the model. 
The first dataset contained EEG recordings from
the Montreal Archive of Sleep Studies (MASS)\cite{OReilly2014},
which are publicly available\cite{MASS2020}.
Retrieval and analysis was approved by the local ethics committee 
at Technische Universität Dresden, Germany (permit number BO-EK-50012021).
All analytical methods were carried out according to the ethical principles of the Declaration of Helsinki\cite{WMA2001} and the recommendations on good clinical practice\cite{EMA2021}.
The second dataset consisted of freely available crowdsourced spindle annotations from the Massive Online Data Annotation
(MODA) study\cite{Lacourse2020}.  Of the 200 individual recordings
contained in MASS, the MODA study selected 180~recordings of C3-LE or C3-M2 EEG channels.
Recordings were split
into a younger (24 years mean age) and an older (62 years mean age) cohort
to make annotations.
Each recording was divided into several 115-second-long segments containing N2
sleep without artifacts.  Ten segments were 
randomly sampled from each recording in a subset of 30
recordings (from 15 younger and 15 older subjects), and three segments were 
randomly sampled 
from each of the
remaining 150 recordings.
After band-pass filtering and downsampling, these segments were exhaustively scored by 
47 certified EEG technicians.
A median of five EEG technicians reviewed any one segment.
Available to use was an expert consensus of spindle annotations formed from the
annotations of the EEG technicians.
The method by which the expert consensus was formed as well
as other details can be found in Lacourse et al. (2020)\cite{Lacourse2020}.

\begin{figure*}[t]
	\centering
	\includegraphics[width=0.65\linewidth]{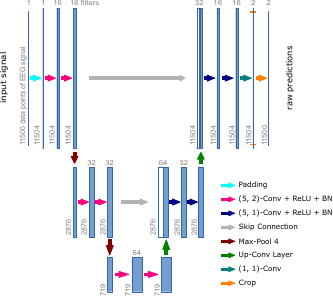}
	\caption{
		Architecture of the SUMO model. 
		Each box represents a feature map with the number of filters given on top.
		The operations conducted are represented by the colored arrows.
		Convolutions are described using the format \textit{(k, d)-Conv} with \textit{k} denoting the kernel size and \textit{d} the dilation size.
		ReLU: rectified linear unit; BN: batch normalization.
		Up-Conv Layer: nearest-neighbor upsampling with factor~4 followed by a (4, 1)-Conv.
	}
	\label{fig:final-architecture}
\end{figure*}

\subsubsection{Signal preprocessing}

We downsampled each EEG segment to 100~Hz using a phase-preserving, 10-th order Butterworth filter (passband 0.3--30 Hz) according to the MODA study.  In addition, we z-transformed each segment (zero mean, unit variance)
before further processing to decrease inter-individual differences
in the recording modalities.

\subsection{Training and test sets}
\label{sec:train-test-split}

We divided the dataset into a training
and a test set (see table~\ref{tab:data:splits}) guided by the following principles.  All segments of any one
individual shall belong to either the training or the test set.  The ten-segment
individuals shall belong to the training set.  The test set shall have equal
numbers of younger and older individuals.  Algorithm A7 shall show an average
performance on the test set segments.

The last principle was added because published results of A7 were obtained on
the whole dataset, for which we were unable to generate an out-of-sample
prediction of our model.
To split off a test set that did not bias A7, we randomly drew 25 test set
candidates from the distribution defined by the rest of the above principles.  We
evaluated A7 on each candidate and determined its F1 score (see section~\ref{sec:detection_evaluation}).  From all candidate
test sets, we chose the one on which A7 reached the median F1 score of the sample
distribution.

\subsection{Model architecture}

The architecture of the model \emph{SUMO} was inspired by U-Net\cite{Ronneberger2015}, which is a
fully-convolutional network. SUMO consisted of a branching encoder making up the left
side of the U followed by a converging decoder making up the right side (see
figure~\ref{fig:final-architecture}).
The encoder compressed temporal features in a single EEG segment into an
increasing number of context channels through several levels of compression.
The decoder expanded the context into temporal information while reducing the
number of context channels in a symmetric number of expansion levels.  After
each compression, a branch was fed into the matching expansive level through
shortcut connections (cf.~figure~\ref{fig:final-architecture}).  The output of the
last layer represented raw per data point probabilities of being part of a spindle or not (no-spindle).
These probabilities 
were either fed into the loss function or underwent further processing to
predict spindle annotations.

Each level of compression and expansion consisted of two composite layers of
convolution\cite{Krizhevsky2012}, rectified linear unit\cite{Nair2010}, and
batch normalization\cite{Ioffe2015}.  The two convolutions had the same number
of channels.  The compression step on the left side of the U was implemented by
max-pooling the composite-layer output.  The expansion step (right side)
was achieved by nearest-neighbor upsampling followed by a convolution mapping to
half the number of input filters.  This half was then supplemented by
concatenating the output activations from the matching contracting step in the
left side of the U.

When predicting spindle annotations, the raw output probabilities in the spindle
and no-spindle channel were temporally smoothed with a moving-average filter and converted to spindle
indicator values by taking the point-wise maximum.  Consecutive positive
indications were then joined to form spindle annotations constituted by a
starting sample and a duration.
The width of the moving-average filter was a hyperparameter that we
optimized (see section~\ref{sec:results}).

\begin{figure*}[t]
	\centering
	\includegraphics[width=\linewidth]{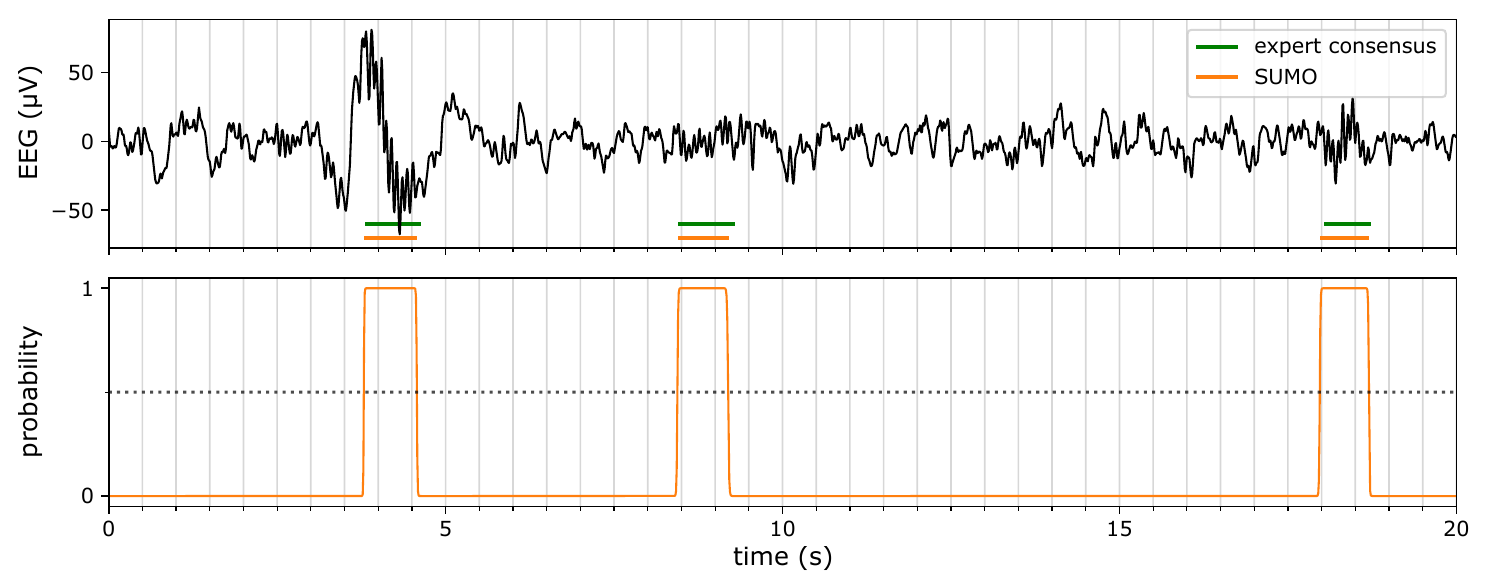}
	\caption{
		Comparison of SUMO predictions with expert consensus on an
		exemplary EEG segment.  Top panel: A 20-second-long segment of
		N2-sleep EEG from the test set shows three spindles marked by
		the expert consensus (green marks).  Our trained SUMO model
		detects all three spindles with about 90\% overlap (orange marks).
		Bottom panel: The predictions are generated by thresholding
		SUMO's output probabilities (orange line) with a threshold of
		$0.5$ (black line).
	}
	\label{fig:spindle-detection-example}
\end{figure*}

\subsection{Training}

Before training, we divided all training segments into training and validation
sets using a six-fold cross validation procedure for hyperparameter optimization
and early stopping.
Similar to the test-training splits, the six folds were split on a by-individual
basis wherein the 10-segment individuals were equally divided across the folds.
This means that the 144 subjects of the training set (see table~\ref{tab:data:splits})
form six folds of 24 subjects each, where a model is trained on five folds
(120 subjects) and evaluated on the remaining fold (24 subjects).
Six separate training sessions were carried out to perform the six-fold cross validation.
For each session, we
initialized the trainable network parameters using Kaiming's procedure of
uniform initialization\cite{KaimingHe2015}.
Feeding the network with minibatches of training examples (12 segments), we
evaluated the output and the expert-consensus annotations using a generalized
dice loss function\cite{Sudre2017}.  The resulting error signals were
backpropagated and parameters were updated using Adam stochastic gradient
descent with parameters $\beta_1=0.9$, $\beta_2=0.999$, $\epsilon=10^{-8}$, and
learning rate $\alpha=0.005$\cite{Kingma2015}.
After processing each sample in the training data once (i.e., after one training
epoch), we computed the network's F1 score (see next section) for the
validation set which was one of the six folds.  If the F1 score did not improve
over the course of 300 training epochs, or if the 800-th training epoch was reached, the training
was stopped.  We then stored the parameters of the best performing models
(one model for each validation fold)
from the training session for further processing.

\subsection{Evaluation of spindle detections}
\label{sec:detection_evaluation}

We evaluated the performance of our model in an \emph{analysis by event} and an
\emph{analysis by subject}. The analysis by event reflects the primary objective
of the model to reliably identify individual sleep spindles.
Following Warby et al. (2014)\cite{Warby2014}, we compared a detected set of
spindles to the expert consensus (EC) on a by-spindle (by-event) basis.  For each EC
spindle, we determined the temporal overlap of the closest detected spindle
relative to the joined duration of both.  If that relative overlap exceeded a
threshold, we counted a true positive (TP).  Else, we counted a false negative (FN).
Each detection that did not sufficiently overlap with an EC spindle was counted
as a false positive (FP).
We note that an alternative to this by-event evaluation would be to evaluate
each data point in a spindle separately. However, such an approach
would put additional weight on long spindles because of temporal
correlations, which we wanted to avoid.

From the evaluations, we computed the recall,
$\text{TP}/(\text{TP}+\text{FN})$, the precision,
$\text{TP}/(\text{TP}+\text{FP})$, and the F1 score,
$2\cdot\text{TP}/(2\cdot\text{TP}+\text{FP}+\text{FN})$, to quantify the
quality of spindle detection.
All these metrics depend on the chosen overlap threshold that determines TP, FN,
and FP. Thus, as a parameter-free metric, we also determined \Fmean{} as the
integral of F1 scores over overlap thresholds.
Finally, for a more detailed \emph{by-event} analysis, we studied SUMO's performance stratified by age cohorts.

In our analysis by subject, we investigated the usefulness of the predictions
for downstream measures such as average spindle characteristics which are of
interest to sleep researchers. From the predicted spindles, we determined
average spindle duration and spindle density per subject. We studied how
much these characteristics were correlated with those derived from the expert
consensus.

\section{Results}
\label{sec:results}

\begin{table*}
	\centering
        \scalebox{0.85}{
	\begin{tabular}{@{}llccccccccc@{}}
		\toprule
		& & \multicolumn{3}{c}{recall} & \multicolumn{3}{c}{precision} & \multicolumn{3}{c}{F1 score} \\
		\cmidrule(l){3-5} \cmidrule(l){6-8} \cmidrule(l){9-11}
		detector & data & both & younger & older & both & younger & older & both & younger & older \\
		\midrule
		mean expert & all & 0.72 & 0.76 & 0.66 & 0.78 & 0.81 & 0.74 & 0.72 & 0.76 & 0.65 \\
		A7\cite{Lacourse2019} & all & 0.73 & 0.75 & 0.70 & 0.71 & 0.73 & 0.69 & 0.72 & 0.74 & 0.70 \\
		\midrule
		A7\cite{Lacourse2019} & test & 0.76 & 0.78 & \textbf{0.73} & 0.70 & 0.70 & 0.69 & 0.73 & 0.74 & 0.71 \\
		SUMO & test & \textbf{0.79} & \textbf{0.82} & \textbf{0.73} & \textbf{0.85} & \textbf{0.85} & \textbf{0.85} & \textbf{0.82} & \textbf{0.84} & \textbf{0.79} \\
		\midrule
		SUMO & val & 0.81 (0.01) & 0.85 (0.02) & 0.75 (0.01) & 0.83 (0.02) & 0.83 (0.03) & 0.83 (0.01) & 0.82 (0.01) & 0.84 (0.02) & 0.79 (0.01) \\
		SUMO & train & 0.82 (0.02) & 0.85 (0.02) & 0.76 (0.03) & 0.83 (0.02) & 0.83 (0.02) & 0.83 (0.02) & 0.82 (0.00) & 0.84 (0.01) & 0.79 (0.01) \\
		\bottomrule
	\end{tabular}}
	\caption{
		Recall, precision, and F1 score of mean expert, A7, and SUMO
		in younger and older individuals.  Our best SUMO model
		numerically exceeds the mean-expert and A7 scores in all
		categories.  The test set performance of A7 exceeds its values
		on the full dataset except for the precision metric.  
		All metrics were computed with a 20\% overlap
		threshold.  Numbers in parentheses denote standard deviations
		across the folds of the 6-fold cross validation.		
		The individual experts were evaluated against the
		leave-one-out expert consensus.
		Maximum values obtained on the test set are shown in bold.
	}
	\label{tab:f1-scores-moda}
\end{table*}

\subsection{Hyperparameter optimization} 

We explored how accurate SUMO could predict the expert
consensus depending on the architecture of the model, i.e. the number of
levels, the maximal receptive field, the number of channels,
and the width of the moving-average filter applied to the net's raw output.
We focussed on these parameters hypothesizing their importance to accurately
predict spindles, which show characteristic time scales in their occurrence,
duration, and rhythm.
For a given configuration, we carried out a six-fold cross validation by
training six SUMO models that were independently initialized. 
After one training epoch of $45$ training steps 
(equivalent to the size of the training set), we applied the model in its
current state to the validation set to determine the F1 score dependent on the
overlap threshold.  From this dependence, we computed \Fmean{} as a
parameter-free metric of the model performance.  If exceeding previous \Fmean{}
values, we noted the new value and stored the parameters as the new candidate
for the best model.  After about $300$~training epochs, the F1 score did typically not
improve further and the training stopped.  The training was also stopped after
$800$~training epochs which occurred in about $50$~individual training sessions.
In figure~\ref{fig:spindle-detection-example}, we show an exemplary spindle
prediction of a trained SUMO model applied to a 20-second-long EEG segment together
with the expert-consensus spindles.  In the example, all three spindles were
detected with an overlap of about $90\%$.

After all six training sessions were concluded, we compared the average across
the best \Fmean{} from each fold to \Fmean{} averages obtained for other network
configurations.
We varied the number of levels of the architecture, and within each level
the width of the max-pooling operation to fine tune their receptive fields. We
also changed the model capacity by varying the number of convolutional filters
per level.
We varied the number of levels of the model between 2 (\Fmean{}$=0.599\pm0.017$)
and 5 (\Fmean{}$=0.635\pm0.015$), and found a shallow optimum of
\Fmean{}$=0.638\pm0.021$ for an architecture consisting of 3~levels.  Here, max
pooling operations went across 4, 4, 2, and 2 data points in consecutive levels
of the model.
For 3-level models, we varied the width $w_2$ of the second max pooling while
keeping the first max-pooling width $w_1$ at $4$.  Between $w_2=1$ (receptive
field of $1.48$~seconds), where we found \Fmean{}$=0.611\pm0.027$, and $w_2=8$
($6.24$~seconds, \Fmean{}$=0.633\pm0.027$), we found a shallow optimum of
\Fmean{}$=0.644\pm0.015$ at $w_2=4$ (i.e.~$3.52$~seconds).
When doubling the number of convolutional filters in the 5-level model (number
of filters per level: 32, 64, 128, 256, 512), we observed
\Fmean{}$=0.629\pm0.022$ compared to \Fmean{}$=0.635\pm0.015$ without doubling
(filters per level: 16, 32, 64, 128, 256). We did not observe large variations
in \Fmean{} when further increasing the number of convolutional filters and thus
used the number of filters as depicted in figure~\ref{fig:final-architecture}.
Finally, we optimized the width of the moving-average filter searching in a range
between $10$ and $200$ data points ($0.1$-$2$~seconds).  We found a shallow
optimum at $42$ data points where we reached \Fmean{}$=0.646$.

In total, we trained about $100$ such configurations and found their F1 scores
to be remarkably stable across folds.  We thus determined the optimal model
configuration within our search space to consist of three levels, each one
pooling across $4$ time points.
The raw output was further processed with a $0.42$-second wide moving-average
filter before thresholding in order to determine spindle predictions.

\subsection{Analysis by event}

We then went on to compare the performance of the SUMO model to algorithm A7
and the expert consensus (see results in table~\ref{tab:f1-scores-moda}). First,
we extracted recall, precision, and F1 score (for 20\% overlap threshold) of the
average expert and A7 from Lacourse et al. (2020)\cite{Lacourse2020},
supplementary table~2. We also computed these metrics for A7 on the hold-out
test set.
Comparing A7 performance on the test and the whole dataset, we found that all
test set metrics were equal to, or numerically exceeded the metrics obtained on
the whole dataset.

We also evaluated SUMO in its best-performing configuration on the test set
and found that its F1 score exceeded A7 by $10$ base points on the younger cohort,
and by $8$ base points on the older cohort.
By comparing all SUMO metrics from the test set to those on the validation and
training set (evaluated for the respective fold), we found no signs of
overfitting across all three sets.

\begin{figure*}[t]
	\centering
	\includegraphics[width=0.7\linewidth]{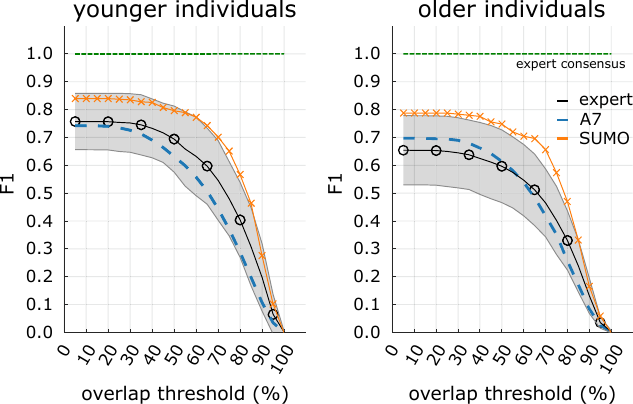}
	\caption{
		Overlap threshold dependency of F1 scores computed against the
		expert consensus.
		The F1 scores of algorithmic (A7: blue dashed line, SUMO:
		orange crosses) and expert (mean: black circles, standard
		deviation: gray area) detections decrease monotonically with
		the overlap threshold for younger (left panel) and older (right
		panel) individuals.
		At all thresholds, SUMO shows a higher F1 score compared
		to A7 as well as most individual experts.
		SUMO F1 scores were computed on the test set, which we designed to
		mimic the whole dataset with respect to the A7 performance.  Expert
		and A7 F1 scores were reproduced from figure~2 in Lacourse et al.
		(2020)\cite{Lacourse2020}, licensed under CC~BY~4.0 International\cite{CCby40}.  As explained therein, the F1 scores of A7
		were computed across the whole dataset, whereas F1 scores for
		individual experts were computed across the scored subset, and against
		all remaining experts.  For comparison, the green dashed line indicates
		the expert consensus which serves as the ground truth.
	}
	\label{fig:f1-scores-moda}
\end{figure*}

We analyzed how our best model compared with average-expert and A7 detections
for overlap thresholds different from 20\%.  We extracted the comparative
performance values from figure~2 in Lacourse et al. (2020)\cite{Lacourse2020}
and overlaid the F1 scores from our best model on the test set.  As shown in
figure~\ref{fig:f1-scores-moda}, we found that our F1 scores were higher compared
to the average expert and A7 for all overlap thresholds.  In the case of the
older individuals, our F1 score was consistently one standard deviation better
(i.e., above the gray area in figure~\ref{fig:f1-scores-moda}) than the average
expert except for overlap thresholds larger than 90\%.

\subsection{Analysis by subject}

Many sleep studies identify sleep spindles to derive characteristics such as the
number of spindles per minute (spindle density), or the average spindle
duration.  We tested how these characteristics derived from automatic spindle
annotations compared to the expert consensus.

\begin{figure*}[t]
	\centering
	\includegraphics[width=\linewidth]{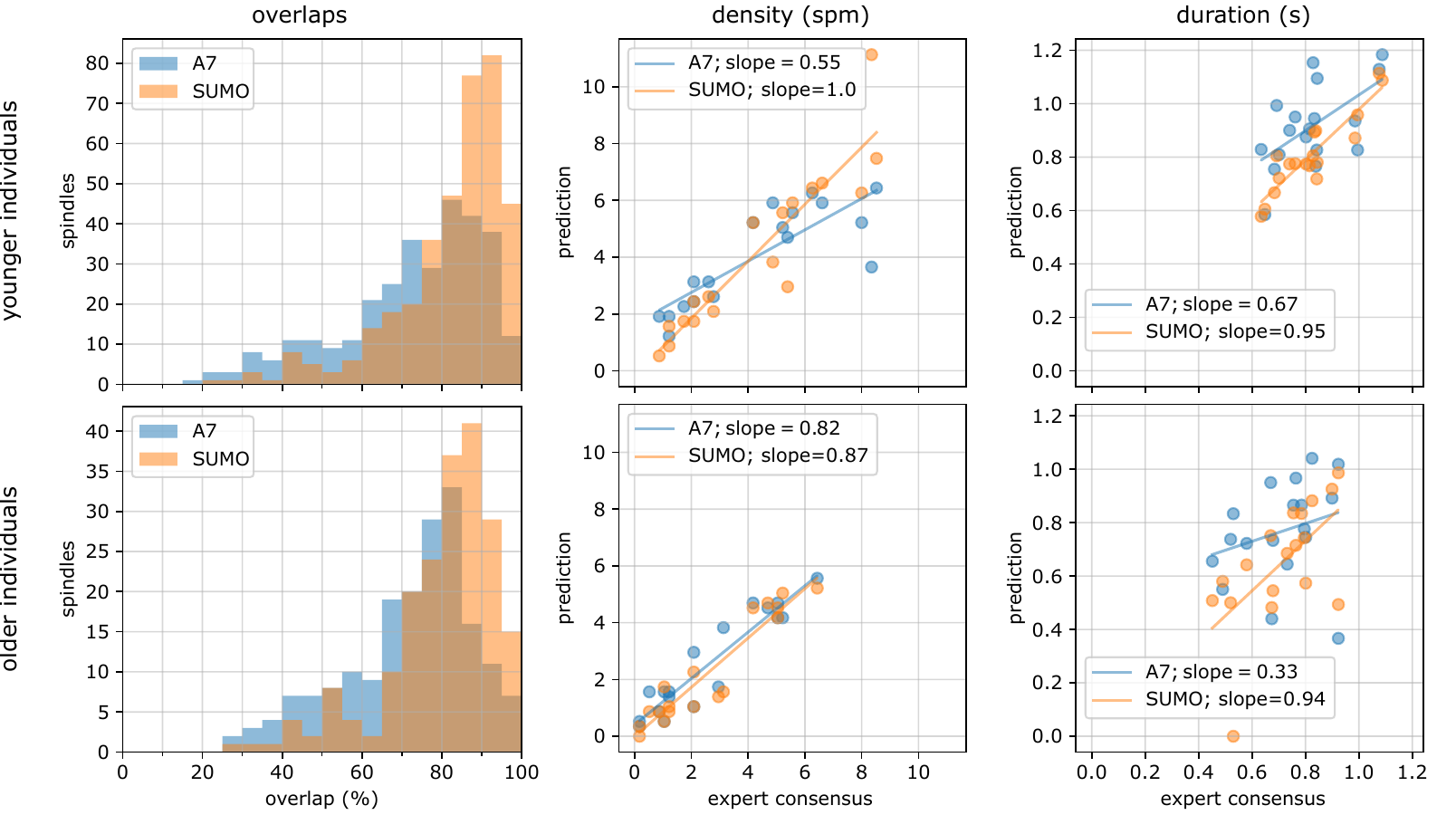}
	\caption{
		Overlap distribution, and spindle density and duration compared
		with expert consensus.  Left panels: The frequency distribution
		of spindle predictions at specific overlaps computed for all
		spindles in the test set is more tightly peaked for SUMO
		(orange) compared with A7 (blue).  The peak of the SUMO model is
		also shifted towards higher overlap percentages.  Middle
		panels: Correlation analysis between per-individual spindle
		density from the expert consensus and predictions (SUMO, A7)
		indicate a flattened slope in younger individuals for A7
		clearly deviating from $1$.  Right panels: Similarly comparing
                spindle durations, A7 also shows a flattened slope.
	}
	\label{fig:spindle-analysis}
\end{figure*}

We determined spindle density and duration for each individual in the test set
from the expert consensus, A7, and the SUMO model presented
in the previous section.  We then analyzed their linear dependency by fitting a
linear model to the relationship of algorithm-derived and consensus
characteristics and noted the slope $m$.  This slope should be $1$ to accurately
indicate changes of the given spindle characteristic caused by an intervention,
for example.
As a measure of precision of the predicted spindle characteristic, we determined
the level of correlation using Pearson's r.  Resultant r-values were compared
across the two algorithms after applying Fisher's z-transform\cite{Fisher1921}.
The results of our analyses are shown in figure~\ref{fig:spindle-analysis} 
(middle and right column)
and table~\ref{tab:moda-correlation}.

\begin{table}[t]
	\centering
	\begin{tabular}{@{}lcccc@{}}
		\toprule
		& \multicolumn{2}{c}{density} & \multicolumn{2}{c}{duration} \\
		\cmidrule(l){2-3} \cmidrule(l){4-5}
		method & younger & older & younger & older \\
		\midrule
		A7\cite{Lacourse2019} & 0.68 & 0.88 & 0.35 & 0.07 \\
		SUMO & 0.84 & 0.89 & 0.82 & 0.37 \\
		\bottomrule
	\end{tabular}
	\caption{
		Correlation coefficient $r^2$ between spindle properties derived by
		automatic methods (A7 or SUMO) and the consensus of the expert group.
		Spindle density (spindles per minute) and average spindle
		duration were determined for each subject of the test set based on
		spindles identified by A7, SUMO, and the expert consensus. For each age
		cohort (younger, older), correlation coefficients between spindle
		properties based on the expert consensus and the automatic methods
		were determined.
	}
	\label{tab:moda-correlation}
\end{table}

\subsubsection{Spindle density}

The expert consensus showed a bimodal distribution of spindle densities with
modes at about $1$ and $5.25$ spindles per minute.
Both algorithm-derived spindle densities showed a high level of correlation with
the expert consensus, with A7 showing $r^2=0.68$ for younger and $r^2=0.88$ for
older participants, and SUMO showing  $r^2=0.84$ and $r^2=0.89$.
When comparing the z-transformed r-values, we found no statistically significant
difference at $\alpha=0.05$.

The slope of A7-derived spindle densities deviated numerically more from $1$ than
SUMO-derived densities, especially for younger individuals with $m_{A7}=0.55$.

\subsubsection{Spindle duration}

In the analysis of spindle duration, we found much weaker correlations compared
to the density results presented above.  A7 showed $r^2=0.35$ for younger and
$r^2=0.07$ for older participants, and SUMO showed $r^2=0.82$ and $r^2=0.37$.
A statistical comparison (based on Fisher's z-transform) gave some evidence that
SUMO derived durations correlated stronger than A7 for younger individuals
($p=0.02$), but not for older ones ($p=0.23$).

When comparing the slope, we found that the A7 fit deviated markedly from an
optimal slope showing $m=0.67$ for younger and $m=0.33$ for older individuals.
In both cases, the SUMO spindle durations showed a slope of about $m=0.95$.

\section{Discussion}
\label{sec:discussion}

We trained the U-Net-type neural network SUMO on the MODA dataset to reliably
identify sleep spindles in the electroencephalographic recordings in a test set
stratified into younger (24 years mean age) and older (62 years mean age)
individuals.
Our hyperparameter tuning optimized \Fmean{}, the integral of F1 scores over
overlap thresholds, computed from the model detections and the
expert-consensus spindles.
We compared the performance metrics \Fmean{}, as well as precision, recall, and
F1 scores for a range of overlap thresholds of our trained model with the A7
algorithm as well as individual experts reported in Lacourse et
al. (2020)\cite{Lacourse2020}.
Note that we did not evaluate the individual expert performance on the test set.
Furthermore, we compared by-subject estimates of clinically-important spindle
metrics (spindle density and average duration) derived by SUMO and A7 to
those determined from the expert-consensus spindles using correlation analysis.

Spindles detected by our model showed, on average, larger overlap with those
of the expert consensus than A7 (see figure~\ref{fig:spindle-analysis}, left
column).
Better agreement with expert-consensus spindles was also visible in F1 scores
which started to decrease at larger overlap thresholds for SUMO compared to A7
(see figure~\ref{fig:f1-scores-moda}).
In line with previous
studies\cite{Warby2014,Lacourse2019,Lacourse2020,Tapia2020}, we evaluated the
performance of our model for an overlap threshold of $20\%$.
SUMO achieved an F1 score of $0.82$ for the subjects in out-of-sample data (test
set, both cohorts, see table~\ref{tab:f1-scores-moda}, third-last column).
This surpassed the A7 algorithm ($0.73$) on the test set and the average expert
($0.72$) on all data.
Our model's performance is comparable to a limited extent with those reported
for other deep learning models at 20\% overlap which were, however, trained on
different datasets (with different age distributions) and predominantly based on
annotations of individual experts.
Reported F1 scores for models evaluated on datasets of young individuals were
comparable to SUMO's F1 score of $0.84$ for the young cohort (F1 scores averaged
over individual experts: $0.83$ (RED-CWT and RED-Time)\cite{Tapia2020}, $0.83$
(SpindleU-Net)\cite{You2021}, $0.83$ (SpindleNet)\cite{Tapia2020}, $0.80$
(DOSED)\cite{Tapia2020}).
These observations may suggest that the accurateness of automated spindle
detection has approached a limit imposed by non-negligible inter- and
intra-rater reliability of experts who annotated the data.
They also provide evidence of the superiority of deep learning
methods compared to classical feature-based approaches such as A7, which
achieved an F1 score of $0.74$ for the younger cohort (see
table~\ref{tab:f1-scores-moda}).
When evaluated on datasets of older individuals, reported performance of other
deep learning models was lower than SUMO's F1 score of $0.79$ for the older
cohort ($0.74$ (SpindleU-Net)\cite{You2021}, $\text{F1}<0.60$
(DOSED)\cite{Chambon2019}).
We note, however, that datasets of older individuals often include patients with
various sleep pathologies, introducing additional factors that may limit
comparative assessment of results.
Nevertheless, the trend towards decreased detection performance was also visible
for A7 (F1 score of $0.71$, see table~\ref{tab:f1-scores-moda}).
The difference in detection performance between age groups may be explained by
spindle amplitude and duration.
Both are known to decrease with age\cite{Nicolas2001,Crowley2002}, which can
make it more difficult for human scorers and automated methods to delineate
spindles from the surrounding background signals in older
individuals\cite{Wendt2015}.
From a clinical perspective, however, we consider the reliable detection of sleep
spindles in older subjects to be of particular importance as sleep disorders
become more frequent with age.
The availability of large datasets with expert-consensus annotations may prove
particularly valuable for the development of spindle detectors and may explain
SUMO's superior performance for this age group.

Detecting individual spindles is only of indirect importance in the evaluation
of a clinical study, for example: to evaluate clinical outcomes, statistical
characteristics of spindles are compared across subjects, raising the question
of how imperfect detection algorithms reflect on such statistical estimates.
In our by-subject analysis of the density and mean duration of spindles, we
evaluated the correlation as well as the slope (from a linear model) between
consensus and automated estimates by A7 and SUMO within age groups.
Spindle densities computed from the expert consensus showed high levels of
correlation with those estimated from A7 ($r^2 \geq 0.68$) and SUMO ($r^2 \geq
0.84$, see table~\ref{tab:moda-correlation}).  Spindle durations, on the other
hand, correlated strongly for SUMO ($r^2 = 0.82$) in the younger cohort, but
only moderately for A7 ($r^2=0.35$), whereas for the older cohort both methods
yielded low correlations ($0.37$ and $0.07$, respectively).
Slopes of the linear models were close to one for SUMO ($m\geq 0.94$) for all
spindle characteristics and age cohorts except for the spindle density of the
younger cohort ($m=0.87$). All slopes for A7 were lower than for SUMO and
achieved the largest value for the spindle density of the older cohort
($m=0.82$) and the lowest for the spindle duration of the older cohort
($m=0.33$).
The results indicate that the spindle density can be estimated with higher
accuracy compared to the mean spindle duration, and that A7 yields a biased
estimator of these quantities.  More complex spindle statistics may show
additional systematic errors, and we recommend that the accurate automatic
estimation of specific metrics should be verified before relying on their
estimates.
We speculate that spindle duration may sensitively depend on large overlap
thresholds between predicted and actual spindles in order for predicted and
actual durations to become similar. As the overlap threshold increased in both
age cohorts, the F1 score decreased for A7 and SUMO, albeit more rapidly for A7
than for SUMO and to lower values for the older cohort than the younger cohort
(see figure~\ref{fig:f1-scores-moda}). This may explain the low levels of
correlation of A7 or SUMO derived spindle duration for the older cohort,
calling for methods that can more accurately identify sleep spindles at large
overlap thresholds.
Thus, we recommend future studies to evaluate spindle detectors at larger overlap
thresholds (e.g., at 40\%).

Our model has low model complexity and few hyperparameters since all layers are
fully convolutional.
Small model complexity allowed the model to be trained in a data-efficient way
which prevents overfitting on small to medium-sized datasets such as MODA and
eliminates the need for sophisticated regularization strategies (e.g., dropout).
Indeed, we did not find evidence of overfitting since the F1 score obtained on
the training set did not exceed the ones obtained on the validation and test
sets ($0.82$, see third-last column in table~\ref{tab:f1-scores-moda}).
Various architectural changes (e.g., number of convolutional filters, the number
of levels of the U, or the receptive fields of neurons of the last levels) did
not sensitively affect the accurateness of spindle detection.
While increasing the number of U-Net levels above three did not change F1
scores, increasing the receptive field of neurons (i.e. the length of an EEG
segment a neuron has access to) of the third level above 3.5 seconds led to a
slight decrease of F1 scores.
This may indicate that particular time scales are relevant for the model to yield
reliable spindle predictions.
As spindle events usually show a duration between 0.5 and 1
second\cite{Lacourse2020}, time scales larger than 3 to 4 seconds may not be
informative for the model to reliably delineate spindle events from surrounding
EEG activity.
Thus, deeper levels of the U that integrate larger time scales may be ignored by
the model.

The present study is limited in several respects.
As previous methods for automated spindle detection, our model was trained to
detect spindles in univariate time series (channel C3).
Such a univariate approach does not localize sleep spindles (scalp
localization\cite{CoppieterstWallant2016}), a task desirable when studying the
topographical distribution of spindles.
While in principle our model can be used repeatedly to identify spindles in
other channels, it remains to be investigated how accurate such detections are.
Accurate localization of spindles could be achieved by multivariate models that
would base their detections on all EEG channels.
To develop and assess such models, we consider the creation of datasets with
expert-consensus annotations for all EEG channels as the most important
prerequisite.
Another limitation shared with previous work\cite{Lacourse2019,Tapia2020} is
also related to the data used to train our model.
Data were selected from N2 sleep stages of healthy individuals and were free
from EEG artifacts.
Ordinary EEG recordings will not meet these criteria, and spindles are also
studied in N3 sleep.
Models that accurately identify sleep spindles in N2 and N3 sleep and account
for artifacts could be developed as datasets with corresponding expert-consensus
annotations become available.
Another limitation is related to the interpretation
of expert-consensus spindles as objective truth.
It is not clear whether spindles detected by models but not identified by the
expert consensus nevertheless are mechanistically identical or may even reflect
other biologically meaningful processes\cite{Lacourse2020}.
Furthermore, if an individual expert identified ``true'' spindles much more
accurately than other experts, the expert consensus would be of lesser quality
compared to this best individual expert.
Since our model approximates an expert consensus, detected spindles are only
objective in the sense that many experts would agree on them.
Finally, we consider comparisons with previously published deep learning models
on the MODA dataset as informative for future model development.
To reduce difficulties for such comparative assessments in future studies, we
publish our model together with detailed instructions\cite{Kaulen2022a}.

Our results underline the potential of deep learning models to automate the
cumbersome task of spindle detection with super-human performance.
We consider the availability of large datasets with expert-consensus annotations
as key driver for the progress of methods towards objective and reliable spindle
detection.
We are confident that automatic spindle detection will continue to improve,
saving staff and sleep researchers time and facilitating large-scale sleep
studies.

\section*{Data availability}

The MODA dataset analyzed in the current study is described in Lacourse et
al. (2020)\cite{Lacourse2020} and is publicly available on the Open Science Framework
(OSF)\cite{Yetton2020}. The MASS dataset\cite{OReilly2014}, which contains the EEG 
recordings, is publicly available and can be obtained from the Montreal Archive
of Sleep Studies web page\cite{MASS2020}. The source code of the SUMO model is available on GitHub\cite{Kaulen2022a}.

\section*{Acknowledgements}

We are grateful to M. Reißel and M. Grajewski for providing us with computing
resources.

\section*{Author contributions}

L.K. conducted the deep learning experiments, analyzed the results, and created
the figures; J.T.C.S and S.B. conceived the deep learning experiments; J.S. and
P.R. helped supervising the project and interpreting the results; S.B. and
J.T.C.S. and L.K. wrote the first draft of the manuscript; all authors reviewed
the manuscript.

\section*{Competing interests}

The authors declare no competing interests.


\begin{thebibliography}{10}
  \urlstyle{rm}
  \expandafter\ifx\csname url\endcsname\relax
    \def\url#1{\texttt{#1}}\fi
  \expandafter\ifx\csname urlprefix\endcsname\relax\def\urlprefix{URL }\fi
  \expandafter\ifx\csname doiprefix\endcsname\relax\def\doiprefix{DOI: }\fi
  \providecommand{\bibinfo}[2]{#2}
  \providecommand{\eprint}[2][]{\href{#1}{#2}}
  
  \bibitem{Fernandez2020}
  \bibinfo{author}{Fernandez, L. M.~J.} \& \bibinfo{author}{L\"{u}thi, A.}
  \newblock \bibinfo{journal}{\bibinfo{title}{Sleep spindles: Mechanisms and
    functions}}.
  \newblock {\emph{\JournalTitle{Physiol. Rev.}}} \textbf{\bibinfo{volume}{100}},
    \bibinfo{pages}{805--868},
    \doiprefix\href{http://doi.org/10.1152/physrev.00042.2018}{10.1152/physrev.00042.2018}
    (\bibinfo{year}{2020}).
  
  \bibitem{Schabus2006}
  \bibinfo{author}{Schabus, M.} \emph{et~al.}
  \newblock \bibinfo{journal}{\bibinfo{title}{Sleep spindle-related activity in
    the human {EEG} and its relation to general cognitive and learning
    abilities}}.
  \newblock {\emph{\JournalTitle{Eur. J. Neurosci.}}}
    \textbf{\bibinfo{volume}{23}}, \bibinfo{pages}{1738--1746},
    \doiprefix\href{http://doi.org/10.1111/j.1460-9568.2006.04694.x}{10.1111/j.1460-9568.2006.04694.x}
    (\bibinfo{year}{2006}).
  
  \bibitem{Fogel2011}
  \bibinfo{author}{Fogel, S.~M.} \& \bibinfo{author}{Smith, C.~T.}
  \newblock \bibinfo{journal}{\bibinfo{title}{The function of the sleep spindle:
    A physiological index of intelligence and a mechanism for sleep-dependent
    memory consolidation}}.
  \newblock {\emph{\JournalTitle{Neurosci. Biobehav. Rev.}}}
    \textbf{\bibinfo{volume}{35}}, \bibinfo{pages}{1154--1165},
    \doiprefix\href{http://doi.org/10.1016/j.neubiorev.2010.12.003}{10.1016/j.neubiorev.2010.12.003}
    (\bibinfo{year}{2011}).
  
  \bibitem{Rasch2013}
  \bibinfo{author}{Rasch, B.} \& \bibinfo{author}{Born, J.}
  \newblock \bibinfo{journal}{\bibinfo{title}{About sleep's role in memory}}.
  \newblock {\emph{\JournalTitle{Physiol. Rev.}}} \textbf{\bibinfo{volume}{93}},
    \bibinfo{pages}{681--766},
    \doiprefix\href{http://doi.org/10.1152/physrev.00032.2012}{10.1152/physrev.00032.2012}
    (\bibinfo{year}{2013}).
  
  \bibitem{Lustenberger2015}
  \bibinfo{author}{Lustenberger, C.}, \bibinfo{author}{Wehrle, F.},
    \bibinfo{author}{T\"{u}shaus, L.}, \bibinfo{author}{Achermann, P.} \&
    \bibinfo{author}{Huber, R.}
  \newblock \bibinfo{journal}{\bibinfo{title}{The multidimensional aspects of
    sleep spindles and their relationship to word-pair memory consolidation}}.
  \newblock {\emph{\JournalTitle{Sleep}}} \textbf{\bibinfo{volume}{38}},
    \bibinfo{pages}{1093--1103},
    \doiprefix\href{http://doi.org/10.5665/sleep.4820}{10.5665/sleep.4820}
    (\bibinfo{year}{2015}).
  
  \bibitem{Normand2016}
  \bibinfo{author}{Normand, M.-P.}, \bibinfo{author}{St-Hilaire, P.} \&
    \bibinfo{author}{Bastien, C.~H.}
  \newblock \bibinfo{journal}{\bibinfo{title}{Sleep spindles characteristics in
    insomnia sufferers and their relationship with sleep misperception}}.
  \newblock {\emph{\JournalTitle{Neural Plast.}}}
    \textbf{\bibinfo{volume}{2016}}, \bibinfo{pages}{1--10},
    \doiprefix\href{http://doi.org/10.1155/2016/6413473}{10.1155/2016/6413473}
    (\bibinfo{year}{2016}).
  
  \bibitem{Ferrarelli2007}
  \bibinfo{author}{Ferrarelli, F.} \emph{et~al.}
  \newblock \bibinfo{journal}{\bibinfo{title}{Reduced sleep spindle activity in
    schizophrenia patients}}.
  \newblock {\emph{\JournalTitle{Am. J. Psychiatry}}}
    \textbf{\bibinfo{volume}{164}}, \bibinfo{pages}{483--492},
    \doiprefix\href{http://doi.org/10.1176/ajp.2007.164.3.483}{10.1176/ajp.2007.164.3.483}
    (\bibinfo{year}{2007}).
  
  \bibitem{Wamsley2012}
  \bibinfo{author}{Wamsley, E.~J.} \emph{et~al.}
  \newblock \bibinfo{journal}{\bibinfo{title}{Reduced sleep spindles and spindle
    coherence in schizophrenia: Mechanisms of impaired memory consolidation?}}
  \newblock {\emph{\JournalTitle{Biol. Psychiatry}}}
    \textbf{\bibinfo{volume}{71}}, \bibinfo{pages}{154--161},
    \doiprefix\href{http://doi.org/10.1016/j.biopsych.2011.08.008}{10.1016/j.biopsych.2011.08.008}
    (\bibinfo{year}{2012}).
  
  \bibitem{Manoach2016}
  \bibinfo{author}{Manoach, D.~S.}, \bibinfo{author}{Pan, J.~Q.},
    \bibinfo{author}{Purcell, S.~M.} \& \bibinfo{author}{Stickgold, R.}
  \newblock \bibinfo{journal}{\bibinfo{title}{Reduced sleep spindles in
    schizophrenia: A treatable endophenotype that links risk genes to impaired
    cognition?}}
  \newblock {\emph{\JournalTitle{Biol. Psychiatry}}}
    \textbf{\bibinfo{volume}{80}}, \bibinfo{pages}{599--608},
    \doiprefix\href{http://doi.org/10.1016/j.biopsych.2015.10.003}{10.1016/j.biopsych.2015.10.003}
    (\bibinfo{year}{2016}).
  
  \bibitem{Limoges2005}
  \bibinfo{author}{Limoges, {\'{E}}.}, \bibinfo{author}{Mottron, L.},
    \bibinfo{author}{Bolduc, C.}, \bibinfo{author}{Berthiaume, C.} \&
    \bibinfo{author}{Godbout, R.}
  \newblock \bibinfo{journal}{\bibinfo{title}{Atypical sleep architecture and the
    autism phenotype}}.
  \newblock {\emph{\JournalTitle{Brain}}} \textbf{\bibinfo{volume}{128}},
    \bibinfo{pages}{1049--1061},
    \doiprefix\href{http://doi.org/10.1093/brain/awh425}{10.1093/brain/awh425}
    (\bibinfo{year}{2005}).
  
  \bibitem{Tessier2015}
  \bibinfo{author}{Tessier, S.} \emph{et~al.}
  \newblock \bibinfo{journal}{\bibinfo{title}{Intelligence measures and stage 2
    sleep in typically-developing and autistic children}}.
  \newblock {\emph{\JournalTitle{Int. J. Psychophysiol.}}}
    \textbf{\bibinfo{volume}{97}}, \bibinfo{pages}{58--65},
    \doiprefix\href{http://doi.org/10.1016/j.ijpsycho.2015.05.003}{10.1016/j.ijpsycho.2015.05.003}
    (\bibinfo{year}{2015}).
  
  \bibitem{Ritter2018}
  \bibinfo{author}{Ritter, P.~S.} \emph{et~al.}
  \newblock \bibinfo{journal}{\bibinfo{title}{Sleep spindles in bipolar disorder
    -- a comparison to healthy control subjects}}.
  \newblock {\emph{\JournalTitle{Acta Psychiat. Scand.}}}
    \textbf{\bibinfo{volume}{138}}, \bibinfo{pages}{163--172},
    \doiprefix\href{http://doi.org/10.1111/acps.12924}{10.1111/acps.12924}
    (\bibinfo{year}{2018}).
  
  \bibitem{Petit2004}
  \bibinfo{author}{Petit, D.}, \bibinfo{author}{Gagnon, J.-F.},
    \bibinfo{author}{Fantini, M.~L.}, \bibinfo{author}{Ferini-Strambi, L.} \&
    \bibinfo{author}{Montplaisir, J.}
  \newblock \bibinfo{journal}{\bibinfo{title}{Sleep and quantitative {EEG} in
    neurodegenerative disorders}}.
  \newblock {\emph{\JournalTitle{J. Psychosom. Res.}}}
    \textbf{\bibinfo{volume}{56}}, \bibinfo{pages}{487--496},
    \doiprefix\href{http://doi.org/10.1016/j.jpsychores.2004.02.001}{10.1016/j.jpsychores.2004.02.001}
    (\bibinfo{year}{2004}).
  
  \bibitem{Weng2020}
  \bibinfo{author}{Weng, Y.-Y.}, \bibinfo{author}{Lei, X.} \&
    \bibinfo{author}{Yu, J.}
  \newblock \bibinfo{journal}{\bibinfo{title}{Sleep spindle abnormalities related
    to alzheimer{\textquotesingle}s disease: a systematic mini-review}}.
  \newblock {\emph{\JournalTitle{Sleep Med.}}} \textbf{\bibinfo{volume}{75}},
    \bibinfo{pages}{37--44},
    \doiprefix\href{http://doi.org/10.1016/j.sleep.2020.07.044}{10.1016/j.sleep.2020.07.044}
    (\bibinfo{year}{2020}).
  
  \bibitem{Berry2018}
  \bibinfo{author}{Berry, R.~B.} \emph{et~al.}
  \newblock \emph{\bibinfo{title}{The {AASM} {M}anual for the {S}coring of
    {S}leep and {A}ssociated {E}vents: {R}ules, {T}erminology and {T}echnical
    {S}pecifications}} (\bibinfo{publisher}{American Academy of Sleep Medicine},
    \bibinfo{address}{Darien, Illinois}, \bibinfo{year}{2018}).
  
  \bibitem{Wendt2015}
  \bibinfo{author}{Wendt, S.~L.} \emph{et~al.}
  \newblock \bibinfo{journal}{\bibinfo{title}{Inter-expert and intra-expert
    reliability in sleep spindle scoring}}.
  \newblock {\emph{\JournalTitle{Clin. Neurophysiol.}}}
    \textbf{\bibinfo{volume}{126}}, \bibinfo{pages}{1548--1556},
    \doiprefix\href{http://doi.org/10.1016/j.clinph.2014.10.158}{10.1016/j.clinph.2014.10.158}
    (\bibinfo{year}{2015}).
  
  \bibitem{Kraemer1979}
  \bibinfo{author}{Kraemer, H.~C.}
  \newblock \bibinfo{journal}{\bibinfo{title}{Ramifications of a population model
    for $\kappa$ as a coefficient of reliability}}.
  \newblock {\emph{\JournalTitle{Psychometrika}}} \textbf{\bibinfo{volume}{44}},
    \bibinfo{pages}{461--472},
    \doiprefix\href{http://doi.org/10.1007/bf02296208}{10.1007/bf02296208}
    (\bibinfo{year}{1979}).
  
  \bibitem{OReilly2014}
  \bibinfo{author}{O{\textquotesingle}Reilly, C.}, \bibinfo{author}{Gosselin,
    N.}, \bibinfo{author}{Carrier, J.} \& \bibinfo{author}{Nielsen, T.}
  \newblock \bibinfo{journal}{\bibinfo{title}{Montreal archive of sleep studies:
    an open-access resource for instrument benchmarking and exploratory
    research}}.
  \newblock {\emph{\JournalTitle{J. Sleep Res.}}} \textbf{\bibinfo{volume}{23}},
    \bibinfo{pages}{628--635},
    \doiprefix\href{http://doi.org/10.1111/jsr.12169}{10.1111/jsr.12169}
    (\bibinfo{year}{2014}).
  
  \bibitem{Devuyst2006}
  \bibinfo{author}{Devuyst, S.} \emph{et~al.}
  \newblock \bibinfo{title}{Automatic sleep spindle detection in patients with
    sleep disorders}.
  \newblock In \emph{\bibinfo{booktitle}{2006 International Conference of the
    {IEEE} Engineering in Medicine and Biology Society}},
    \bibinfo{pages}{3883--3886},
    \doiprefix\href{http://doi.org/10.1109/iembs.2006.259298}{10.1109/iembs.2006.259298}
    (\bibinfo{publisher}{{IEEE}}, \bibinfo{address}{New York, NY, USA},
    \bibinfo{year}{2006}).
  
  \bibitem{Kulkarni2019}
  \bibinfo{author}{Kulkarni, P.~M.} \emph{et~al.}
  \newblock \bibinfo{journal}{\bibinfo{title}{A deep learning approach for
    real-time detection of sleep spindles}}.
  \newblock {\emph{\JournalTitle{J. Neural Eng.}}} \textbf{\bibinfo{volume}{16}},
    \bibinfo{pages}{036004},
    \doiprefix\href{http://doi.org/10.1088/1741-2552/ab0933}{10.1088/1741-2552/ab0933}
    (\bibinfo{year}{2019}).
  
  \bibitem{Lacourse2020}
  \bibinfo{author}{Lacourse, K.}, \bibinfo{author}{Yetton, B.},
    \bibinfo{author}{Mednick, S.} \& \bibinfo{author}{Warby, S.~C.}
  \newblock \bibinfo{journal}{\bibinfo{title}{Massive online data annotation,
    crowdsourcing to generate high quality sleep spindle annotations from {EEG}
    data}}.
  \newblock {\emph{\JournalTitle{Sci. Data}}} \textbf{\bibinfo{volume}{7}},
    \bibinfo{pages}{190},
    \doiprefix\href{http://doi.org/10.1038/s41597-020-0533-4}{10.1038/s41597-020-0533-4}
    (\bibinfo{year}{2020}).
  
  \bibitem{Lacourse2019}
  \bibinfo{author}{Lacourse, K.}, \bibinfo{author}{Delfrate, J.},
    \bibinfo{author}{Beaudry, J.}, \bibinfo{author}{Peppard, P.} \&
    \bibinfo{author}{Warby, S.~C.}
  \newblock \bibinfo{journal}{\bibinfo{title}{A sleep spindle detection algorithm
    that emulates human expert spindle scoring}}.
  \newblock {\emph{\JournalTitle{J. Neurosci. Methods}}}
    \textbf{\bibinfo{volume}{316}}, \bibinfo{pages}{3--11},
    \doiprefix\href{http://doi.org/10.1016/j.jneumeth.2018.08.014}{10.1016/j.jneumeth.2018.08.014}
    (\bibinfo{year}{2019}).
  
  \bibitem{Schimicek1994}
  \bibinfo{author}{Schimicek, P.}, \bibinfo{author}{Zeitlhofer, J.},
    \bibinfo{author}{Anderer, P.} \& \bibinfo{author}{Saletu, B.}
  \newblock \bibinfo{journal}{\bibinfo{title}{Automatic sleep-spindle detection
    procedure: Aspects of reliability and validity}}.
  \newblock {\emph{\JournalTitle{Clin. Electroencephal.}}}
    \textbf{\bibinfo{volume}{25}}, \bibinfo{pages}{26--29},
    \doiprefix\href{http://doi.org/10.1177/155005949402500108}{10.1177/155005949402500108}
    (\bibinfo{year}{1994}).
  
  \bibitem{Moelle2002}
  \bibinfo{author}{M\"{o}lle, M.}, \bibinfo{author}{Marshall, L.},
    \bibinfo{author}{Gais, S.} \& \bibinfo{author}{Born, J.}
  \newblock \bibinfo{journal}{\bibinfo{title}{Grouping of spindle activity during
    slow oscillations in human non-rapid eye movement sleep}}.
  \newblock {\emph{\JournalTitle{J. Neurosci.}}} \textbf{\bibinfo{volume}{22}},
    \bibinfo{pages}{10941--10947},
    \doiprefix\href{http://doi.org/10.1523/jneurosci.22-24-10941.2002}{10.1523/jneurosci.22-24-10941.2002}
    (\bibinfo{year}{2002}).
  
  \bibitem{Martin2013}
  \bibinfo{author}{Martin, N.} \emph{et~al.}
  \newblock \bibinfo{journal}{\bibinfo{title}{Topography of age-related changes
    in sleep spindles}}.
  \newblock {\emph{\JournalTitle{Neurobiol. Aging}}}
    \textbf{\bibinfo{volume}{34}}, \bibinfo{pages}{468--476},
    \doiprefix\href{http://doi.org/10.1016/j.neurobiolaging.2012.05.020}{10.1016/j.neurobiolaging.2012.05.020}
    (\bibinfo{year}{2013}).
  
  \bibitem{Parekh2015}
  \bibinfo{author}{Parekh, A.}, \bibinfo{author}{Selesnick, I.~W.},
    \bibinfo{author}{Rapoport, D.~M.} \& \bibinfo{author}{Ayappa, I.}
  \newblock \bibinfo{journal}{\bibinfo{title}{Detection of k-complexes and sleep
    spindles ({DETOKS}) using sparse optimization}}.
  \newblock {\emph{\JournalTitle{J. Neurosci. Methods}}}
    \textbf{\bibinfo{volume}{251}}, \bibinfo{pages}{37--46},
    \doiprefix\href{http://doi.org/10.1016/j.jneumeth.2015.04.006}{10.1016/j.jneumeth.2015.04.006}
    (\bibinfo{year}{2015}).
  
  \bibitem{Schmidhuber2015}
  \bibinfo{author}{Schmidhuber, J.}
  \newblock \bibinfo{journal}{\bibinfo{title}{Deep learning in neural networks:
    {A}n overview}}.
  \newblock {\emph{\JournalTitle{Neural Networks}}}
    \textbf{\bibinfo{volume}{61}}, \bibinfo{pages}{85--117},
    \doiprefix\href{http://doi.org/10.1016/j.neunet.2014.09.003}{10.1016/j.neunet.2014.09.003}
    (\bibinfo{year}{2015}).
  
  \bibitem{LeCun2015}
  \bibinfo{author}{LeCun, Y.}, \bibinfo{author}{Bengio, Y.} \&
    \bibinfo{author}{Hinton, G.}
  \newblock \bibinfo{journal}{\bibinfo{title}{Deep learning}}.
  \newblock {\emph{\JournalTitle{Nature}}} \textbf{\bibinfo{volume}{521}},
    \bibinfo{pages}{436--444},
    \doiprefix\href{http://doi.org/10.1038/nature14539}{10.1038/nature14539}
    (\bibinfo{year}{2015}).
  
  \bibitem{Goodfellow2016}
  \bibinfo{author}{Goodfellow, I.~J.}, \bibinfo{author}{Bengio, Y.} \&
    \bibinfo{author}{Courville, A.~C.}
  \newblock \emph{\bibinfo{title}{Deep {L}earning}}.
  \newblock Adaptive computation and machine learning (\bibinfo{publisher}{{MIT}
    Press}, \bibinfo{address}{Cambridge, Massachusetts}, \bibinfo{year}{2016}).
  
  \bibitem{Chambon2019}
  \bibinfo{author}{Chambon, S.}, \bibinfo{author}{Thorey, V.},
    \bibinfo{author}{Arnal, P.~J.}, \bibinfo{author}{Mignot, E.} \&
    \bibinfo{author}{Gramfort, A.}
  \newblock \bibinfo{journal}{\bibinfo{title}{{DOSED}: A deep learning approach
    to detect multiple sleep micro-events in {EEG} signal}}.
  \newblock {\emph{\JournalTitle{J. Neurosci. Methods}}}
    \textbf{\bibinfo{volume}{321}}, \bibinfo{pages}{64--78},
    \doiprefix\href{http://doi.org/10.1016/j.jneumeth.2019.03.017}{10.1016/j.jneumeth.2019.03.017}
    (\bibinfo{year}{2019}).
  
  \bibitem{Tapia2020}
  \bibinfo{author}{Tapia, N.~I.} \& \bibinfo{author}{Estevez, P.~A.}
  \newblock \bibinfo{title}{{RED}: Deep recurrent neural networks for sleep {EEG}
    event detection}.
  \newblock In \emph{\bibinfo{booktitle}{Int. Joint Conf. on Neural Networks
    ({IJCNN})}}, \bibinfo{pages}{1--8},
    \doiprefix\href{http://doi.org/10.1109/ijcnn48605.2020.9207719}{10.1109/ijcnn48605.2020.9207719}
    (\bibinfo{publisher}{{IEEE}}, \bibinfo{address}{Glasgow, UK},
    \bibinfo{year}{2020}).
  
  \bibitem{You2021}
  \bibinfo{author}{You, J.}, \bibinfo{author}{Jiang, D.}, \bibinfo{author}{Ma,
    Y.} \& \bibinfo{author}{Wang, Y.}
  \newblock \bibinfo{journal}{\bibinfo{title}{{SpindleU-Net}: An adaptive u-net
    framework for sleep spindle detection in single-channel {EEG}}}.
  \newblock {\emph{\JournalTitle{{IEEE} Trans. Neural Syst. Rehabilitation
    Eng.}}} \textbf{\bibinfo{volume}{29}}, \bibinfo{pages}{1614--1623},
    \doiprefix\href{http://doi.org/10.1109/tnsre.2021.3105443}{10.1109/tnsre.2021.3105443}
    (\bibinfo{year}{2021}).
  
  \bibitem{Kaulen2022a}
  \bibinfo{author}{Kaulen, L.}, \bibinfo{author}{Schwabedal, J. T.~C.} \&
    \bibinfo{author}{Bialonski, S.}
  \newblock \bibinfo{title}{Source code of the model presented in {K}aulen
    \emph{et al.}, ``{A}dvanced sleep spindle identification with neural
    networks''}.
  \newblock \bibinfo{howpublished}{\url{https://github.com/dslaborg/sumo}}
    (\bibinfo{year}{2022}).
  
  \bibitem{Ronneberger2015}
  \bibinfo{author}{Ronneberger, O.}, \bibinfo{author}{Fischer, P.} \&
    \bibinfo{author}{Brox, T.}
  \newblock \bibinfo{title}{{U}-{N}et: Convolutional networks for biomedical
    image segmentation}.
  \newblock In \emph{\bibinfo{booktitle}{Lecture Notes in Computer Science}},
    \bibinfo{pages}{234--241},
    \doiprefix\href{http://doi.org/10.1007/978-3-319-24574-4\_28}{10.1007/978-3-319-24574-4\_28}
    (\bibinfo{publisher}{Springer International Publishing},
    \bibinfo{year}{2015}).
  
  \bibitem{Siddique2021}
  \bibinfo{author}{Siddique, N.}, \bibinfo{author}{Paheding, S.},
    \bibinfo{author}{Elkin, C.~P.} \& \bibinfo{author}{Devabhaktuni, V.}
  \newblock \bibinfo{journal}{\bibinfo{title}{{U-N}et and its variants for
    medical image segmentation: {A} review of theory and applications}}.
  \newblock {\emph{\JournalTitle{{IEEE} {A}ccess}}} \textbf{\bibinfo{volume}{9}},
    \bibinfo{pages}{82031--82057},
    \doiprefix\href{http://doi.org/10.1109/access.2021.3086020}{10.1109/access.2021.3086020}
    (\bibinfo{year}{2021}).
  
  \bibitem{MASS2020}
  \bibinfo{title}{Web page of the {MASS} dataset}.
  \newblock \bibinfo{howpublished}{\url{http://ceams-carsm.ca/mass/}}
    (\bibinfo{year}{2020}).
  
  \bibitem{WMA2001}
  \bibinfo{author}{{World Medical Association}}.
  \newblock \bibinfo{journal}{\bibinfo{title}{World medical association
    {D}eclaration of {H}elsinki. {E}thical principles for medical research
    involving human subjects}}.
  \newblock {\emph{\JournalTitle{Bull. World Health Organ.}}}
    \textbf{\bibinfo{volume}{79}}, \bibinfo{pages}{373--374}
    (\bibinfo{year}{2001}).
  
  \bibitem{EMA2021}
  \bibinfo{author}{{European Medicines Agency}}.
  \newblock \bibinfo{title}{Good clinical practice}.
  \newblock
    \bibinfo{howpublished}{\url{https://www.ema.europa.eu/en/human-regulatory/research-development/compliance/good-clinical-practice}}
    (\bibinfo{year}{2021}).
  
  \bibitem{Krizhevsky2012}
  \bibinfo{author}{Krizhevsky, A.}, \bibinfo{author}{Sutskever, I.} \&
    \bibinfo{author}{Hinton, G.~E.}
  \newblock \bibinfo{title}{Image{N}et classification with deep convolutional
    neural networks}.
  \newblock In \emph{\bibinfo{booktitle}{Advances in Neural Information
    Processing Systems 25, {N}eur{IPS}}}, \bibinfo{pages}{1106--1114}
    (\bibinfo{year}{2012}).
  
  \bibitem{Nair2010}
  \bibinfo{author}{Nair, V.} \& \bibinfo{author}{Hinton, G.~E.}
  \newblock \bibinfo{title}{Rectified linear units improve restricted {B}oltzmann
    machines}.
  \newblock In \emph{\bibinfo{booktitle}{Proc. 27th Int. Conf. Machine Learning,
    {ICML}}}, \bibinfo{pages}{807--814} (\bibinfo{address}{Haifa, Israel},
    \bibinfo{year}{2010}).
  
  \bibitem{Ioffe2015}
  \bibinfo{author}{Ioffe, S.} \& \bibinfo{author}{Szegedy, C.}
  \newblock \bibinfo{title}{Batch {N}ormalization: accelerating deep network
    training by reducing internal covariate shift}.
  \newblock In \emph{\bibinfo{booktitle}{Proc. 32nd Int. Conf. Machine Learning,
    {ICML}}}, \bibinfo{pages}{448--456} (\bibinfo{address}{Lille, France},
    \bibinfo{year}{2015}).
  
  \bibitem{KaimingHe2015}
  \bibinfo{author}{He, K.}, \bibinfo{author}{Zhang, X.}, \bibinfo{author}{Ren,
    S.} \& \bibinfo{author}{Sun, J.}
  \newblock \bibinfo{title}{Delving deep into rectifiers: {S}urpassing
    human-level performance on {ImageNet} classification}.
  \newblock In \emph{\bibinfo{booktitle}{{IEEE} Int. Conf. on Computer Vision,
    {ICCV}}}, \bibinfo{pages}{1026--1034},
    \doiprefix\href{http://doi.org/10.1109/ICCV.2015.123}{10.1109/ICCV.2015.123}
    (\bibinfo{publisher}{{IEEE} Computer Society}, \bibinfo{address}{Santiago,
    Chile}, \bibinfo{year}{2015}).
  
  \bibitem{Sudre2017}
  \bibinfo{author}{Sudre, C.~H.}, \bibinfo{author}{Li, W.},
    \bibinfo{author}{Vercauteren, T.}, \bibinfo{author}{Ourselin, S.} \&
    \bibinfo{author}{Cardoso, M.~J.}
  \newblock \bibinfo{title}{Generalised dice overlap as a deep learning loss
    function for highly unbalanced segmentations}.
  \newblock In \emph{\bibinfo{booktitle}{Deep Learning in Medical Image Analysis
    and Multimodal Learning for Clinical Decision Support}},
    \bibinfo{pages}{240--248},
    \doiprefix\href{http://doi.org/10.1007/978-3-319-67558-9\_28}{10.1007/978-3-319-67558-9\_28}
    (\bibinfo{publisher}{Springer International Publishing},
    \bibinfo{year}{2017}).
  
  \bibitem{Kingma2015}
  \bibinfo{author}{Kingma, D.~P.} \& \bibinfo{author}{Ba, J.}
  \newblock \bibinfo{title}{Adam: a method for stochastic optimization}.
  \newblock In \emph{\bibinfo{booktitle}{3rd Int. Conf. Learning Representations,
    {ICLR}}} (\bibinfo{address}{San Diego, CA, USA}, \bibinfo{year}{2015}).
  \newblock \eprint[https://arxiv.org/abs/1412.6980v9]{1412.6980v9}.
  
  \bibitem{Warby2014}
  \bibinfo{author}{Warby, S.~C.} \emph{et~al.}
  \newblock \bibinfo{journal}{\bibinfo{title}{Sleep-spindle detection:
    crowdsourcing and evaluating performance of experts, non-experts and
    automated methods}}.
  \newblock {\emph{\JournalTitle{Nat. Methods}}} \textbf{\bibinfo{volume}{11}},
    \bibinfo{pages}{385--392},
    \doiprefix\href{http://doi.org/10.1038/nmeth.2855}{10.1038/nmeth.2855}
    (\bibinfo{year}{2014}).
  
  \bibitem{CCby40}
  \bibinfo{title}{{C}reative {C}ommons {A}ttribution 4.0 {I}nternational {P}ublic
    {L}icense}.
  \newblock
    \bibinfo{howpublished}{\url{https://creativecommons.org/licenses/by/4.0/}}.
  
  \bibitem{Fisher1921}
  \bibinfo{author}{Fisher, R.~A.}
  \newblock \bibinfo{journal}{\bibinfo{title}{On the "probable error" of a
    coefficient of correlation deduced from a small sample.}}
  \newblock {\emph{\JournalTitle{Metron}}} \textbf{\bibinfo{volume}{1}},
    \bibinfo{pages}{3--32} (\bibinfo{year}{1921}).
  
  \bibitem{Nicolas2001}
  \bibinfo{author}{Nicolas, A.}, \bibinfo{author}{Petit, D.},
    \bibinfo{author}{Rompr{\'{e}}, S.} \& \bibinfo{author}{Montplaisir, J.}
  \newblock \bibinfo{journal}{\bibinfo{title}{Sleep spindle characteristics in
    healthy subjects of different age groups}}.
  \newblock {\emph{\JournalTitle{Clin. Neurophysiol.}}}
    \textbf{\bibinfo{volume}{112}}, \bibinfo{pages}{521--527},
    \doiprefix\href{http://doi.org/10.1016/s1388-2457(00)00556-3}{10.1016/s1388-2457(00)00556-3}
    (\bibinfo{year}{2001}).
  
  \bibitem{Crowley2002}
  \bibinfo{author}{Crowley, K.}, \bibinfo{author}{Trinder, J.},
    \bibinfo{author}{Kim, Y.}, \bibinfo{author}{Carrington, M.} \&
    \bibinfo{author}{Colrain, I.~M.}
  \newblock \bibinfo{journal}{\bibinfo{title}{The effects of normal aging on
    sleep spindle and {K}-complex production}}.
  \newblock {\emph{\JournalTitle{Clin. Neurophysiol.}}}
    \textbf{\bibinfo{volume}{113}}, \bibinfo{pages}{1615--1622},
    \doiprefix\href{http://doi.org/10.1016/s1388-2457(02)00237-7}{10.1016/s1388-2457(02)00237-7}
    (\bibinfo{year}{2002}).
  
  \bibitem{CoppieterstWallant2016}
  \bibinfo{author}{Coppieters {\textquotesingle}t~Wallant, D.},
    \bibinfo{author}{Maquet, P.} \& \bibinfo{author}{Phillips, C.}
  \newblock \bibinfo{journal}{\bibinfo{title}{Sleep spindles as an electrographic
    element: Description and automatic detection methods}}.
  \newblock {\emph{\JournalTitle{Neural Plast.}}}
    \textbf{\bibinfo{volume}{2016}}, \bibinfo{pages}{1--19},
    \doiprefix\href{http://doi.org/10.1155/2016/6783812}{10.1155/2016/6783812}
    (\bibinfo{year}{2016}).
  
  \bibitem{Yetton2020}
  \bibinfo{author}{Yetton, B.~D.}, \bibinfo{author}{Lacourse, K.},
    \bibinfo{author}{Delfrate, J.}, \bibinfo{author}{Mednick, S.~C.} \&
    \bibinfo{author}{Warby, S.}
  \newblock \bibinfo{title}{The {MODA} sleep spindle dataset: {A} large, open,
    high quality dataset of annotated sleep spindles},
    \doiprefix\href{http://doi.org/10.17605/OSF.IO/8BMA7}{10.17605/OSF.IO/8BMA7}
    (\bibinfo{year}{2020}).
  
  \end{thebibliography}
\end{document}